\documentclass[aps,pra,preprint,superscriptaddress,showkeys,showpacs]{revtex4}
\usepackage[dvips]{graphics}

\newcommand{\be}{\begin{equation}}
\newcommand{\ee}{\end{equation}}
\newcommand{\bea}{\begin{eqnarray}}
\newcommand{\eea}{\end{eqnarray}}

\newcommand{\itDelta}{{\it \Delta}}

\newcommand{\bra}{\langle}
\newcommand{\ket}{\rangle}

\newcommand{\bq}{{\bar q}}

\newcommand{\rS}{{\rm S}}

\begin{document}


\draft

\title{Multifractal PDF analysis for intermittent systems
}

\author{T.~Arimitsu}
\email[T.~Arimitsu: ]{tarimtsu@sakura.cc.tsukuba.ac.jp}
\address{Graduate School of Pure and Applied Sciences, University of Tsukuba}

\author{N.~Arimitsu}
\address{Graduate School of Environment and Information Sciences, 
Yokohama National University}

\date{\today}

\begin{abstract}
The formula for probability density functions (PDFs) has been extended to 
include PDF for energy dissipation rates in addition to other PDFs such as 
for velocity fluctuations, velocity derivatives, fluid particle accelerations, 
energy transfer rates, etc, and it is shown that the formula actually explains 
various PDFs extracted from direct numerical simulations and experiments 
performed in a wind tunnel. It is also shown that the 
formula with appropriate zooming increment corresponding to experimental 
situation gives a new route to obtain the scaling exponents of velocity 
structure function, including intermittency exponent, out of PDFs of 
velocity fluctuations.
\end{abstract}


\maketitle



\section{Introduction}

The quest for an essence of intermittency, i.e., a {\it fundamental process} 
in turbulence, has a long history but still an unsolved problem in physics 
for more than 120 years since about 1880 when the systematic experiments of 
turbulence was started by Reynolds. The theoretical research on the subject 
in fully developed turbulence (we simply call it turbulence in the following 
unless it is confusing) starts with 
Kolmogorov's dimensional analysis (K41) \cite{K41}
based on the assumption of the self-similarity of fluctuating velocity field 
in the inertial range. After Landau's criticism against K41 in about 1944 
and the preliminary research by Heisenberg \cite{Heisenberg48b}, the quest develops, 
mainly, into two directions. One is the {\it dynamical approach}; 
the other is the {\it ensemble approach} \cite{FrischBook95}. 
Within the dynamical approach 
one treats the stochastic Navier-Stokes equation by perturbational methods
whereas within the ensemble approach one performs statistical mechanical 
analysis of turbulence under the assumption that eddies make up energy 
cascade. It has been, gradually, revealed 
\cite{AA,AA1,AA4,AA7,AA8,AA13,AA15}
that, among the ensemble methods, a new theoretical framework named 
{\it multifractal probability density function analysis} (MPDFA) and 
A\&A model within the framework can analyze in a high precision 
the data extracted out from the recent experiments and simulations 
conducted with higher accuracy.
 
Several quantities such as velocity derivatives, fluid particle accelerations 
and energy dissipation rates have some singularities due to the invariance 
of the Navier-Stokes equation for velocity field ${\vec u}({\vec x},t)$ 
in high Reynolds numbers under the scale transformation:
\be
{\vec x} \rightarrow {\vec x}'=\lambda {\vec x},\quad 
{\vec u} \rightarrow {\vec u}'=\lambda^{\alpha/3} {\vec u},\quad 
t \rightarrow t'=\lambda^{1- \alpha/3} t,\quad 
p \rightarrow p'=\lambda^{2\alpha/3} p
\label{scale trans}
\ee
with arbitrary real number $\alpha$ \cite{Frisch-Parisi83}. 
Here, $p$ represents pressure. 
The MPDFA is a statistical mechanical theory of an ensemble providing 
analytical formulae for various probability density functions (PDFs) 
applicable to intermittent systems. It was constructed on the assumption 
that the strengths of the singularities distribute themselves in a 
multifractal way in real physical space. This distribution of 
singularities determines the tail part of PDFs. The parameters 
appeared in the theory are determined, uniquely, by the intermittency 
exponent $\mu$   that represents the strength of intermittency.

On the other hand, observed PDFs should include the effect resulted from 
the term in the Navier-Stokes equation that violates the invariance under the scale 
transformation (the dissipative term). There has been, however, 
no ensemble theory of turbulence taking this effect into account, 
and the situation remained at the stage where almost all the theories 
are just trying to explain observed {\it scaling exponents} 
$\zeta_{m}$ of the $m$th order velocity structure function, i.e., 
the $m$th moment of velocity fluctuations. The MPDFA counts 
this effect as something determining the central part of PDFs narrower 
than its standard deviation. We are assuming that the fat-tail part, 
which the PDFs of intermittent systems took on, is determined by 
the global characteristics of the system, and that the central part of 
PDFs is a reflection of the local nature of constituting eddies.

\section{Formula for A\&A model within MPDFA\label{sec MPDFA}}

The scaling exponents of the velocity structure function within 
A\&A model is given by \cite{AA1,AA4}
\be
\zeta_m = \alpha_0 m/3 
- 2Xm^2/\left[9 \left(1+\sqrt{C_{m/3}} \right)\right]
- \left[1-\log_2 \left(1+\sqrt{C_{m/3}} \right) \right] /(1-q)
\label{zeta}
\ee
with
$
{C}_{m} = 1 + 2 m^2 (1-q) X \ln 2
\label{cal D}
$.
The parameters $\alpha_0$, $X$ and $q$ are introduced through the Tsallis-type distribution function
\be
P^{(n)}(\alpha) d\alpha \propto 
\left\{ 1 - \left[(\alpha - \alpha_0)/\itDelta \alpha \right]^2 
\right\}^{n/(1-q)} d\alpha
\label{Tsallis prob density}
\ee
with 
$
(\itDelta \alpha)^2 = 2X \big/ [(1-q) \ln 2 ]
$
adopted within A\&A model as the probability to find a singularity specified by $\alpha$  within the range 
$
\alpha \sim \alpha + d \alpha
$
for large $n$, and are determined self-consistently as functions of $\mu$ through the relation
$
\mu = 2 - \zeta_6
$.
This determines the distribution of the arbitrary real number $\alpha$ 
appeared in the scale transformation (\ref{scale trans}).

When the scaling exponents are given by numerical or ordinary experiments, 
we analyze them with the formula (\ref{zeta}) to determine the value $\mu$. 
When they are not given experimentally, we have another route to determine 
its value with the help of the observed PDFs. The latter new route 
is provided first in this paper within MPDFA.


We list here a unified formula for PDFs $\Pi_{\phi}^{(n)} (x_n)$  within A\&A model of MPDFA. The contribution of singularities to PDF is taken into account by
\be
\Pi^{(n)}_{\phi,\rS}(\vert x_n \vert) d(\vert x_n \vert) \propto P^{(n)}(\alpha) d \alpha
\label{singular portion}
\ee
with the transformation of variables
\bea
\vert x_n \vert = \delta x_{n}/\delta x_{0} = \delta_{n}^{\alpha \phi/3}, \quad
\delta_{n} = \ell_{n}/\ell_{0} = 2^{-n}, \quad
\delta x_n = \vert x(\bullet + \ell_n) - x(\bullet) \vert.
\label{trans vari}
\eea
Here, $x$ represents an observable quantity such as a component of fluid 
velocity field $\vec u$, pressure $p$, etc., and $n$ does a number of 
multifractal steps whose increment $\it\Delta n$ gives the {\it zooming 
increment} that should correspond to the process how experimentalists 
extracted PDFs by changing the consecutive distances $r=\ell_{n}$ 
between two observing points, say $r'$ and $r$, i.e., with an appropriate $\mu$ 
the correct zooming increment $\it\Delta n = n' - n$ is provided by
$
\it\Delta n = -\log_2(r'/r)
$
with 
$
n= -\log_2(r/\eta) + \log_2(\ell_0/\eta)
$
where $\eta$ is the Kolmogolov length. Note that $\ell_{0}$ is a 
reference length that is not necessarily equal to the integral 
length $\ell_{\rm in}$ in general \cite{AA13}.

The {\it tail part} $(\vert \xi_{n} \vert > \xi_{n}^{*})$ of the PDF 
for variable  $\xi_{n}$ defined in the ranges ($-\infty$, $\infty$) 
and (0, $\infty$) is given by
\be
\hat{\Pi}_{\phi,\mbox{tl}}^{(n)}(\xi_n) d\xi_n=
\hat{\Pi}_{\phi,S}^{(n)}(x_n)  dx_n 
\propto
\frac{\bar{\xi}_n}{\vert \xi_n \vert}
\left[1-\frac{1-q}{n}\frac{(3\ln \vert \xi_n/\xi_{n,0})^2}{2 \phi^2 X 
\vert \ln \delta_n \vert}\right]^{n/(1-q)} d \xi_n.
\label{PDF tl}
\ee
The {\it center part} $(\vert \xi_n \vert < \xi_n^{*})$ of 
the PDF for variable $\xi_n$ defined in the range ($-\infty$, $\infty$) is given by
\be
\hat{\Pi}_{\phi,\mbox{cr}}^{(n)}(\xi_n) \propto
\left\{1-(1-q')\frac{\phi+3f'(\alpha^*)}{2\phi}
\left[ \left(\frac{\xi_n}{\xi_n^*}\right)^2 -1 \right] \right\}^{1/(1-q')}
\label{PDF cra}
\ee
whereas for variable $\xi_n$ defined in the range (0, $\infty$) by
\be
\hat{\Pi}_{\phi,\mbox{cr}}^{(n)}(\xi_n) \propto
\left(\frac{\xi_n}{\xi_n^*}\right)^{\theta-1} 
\left\{1-(1-q')\frac{\phi\theta+3f'(\alpha^*)}{2\phi}
\left[ \left(\frac{\xi_n}{\xi_n^*}\right)^2 -1 \right] \right\}^{1/(1-q')}.
\label{PDF crb}
\ee
The variable $\xi_n$ is the scaled variable related to an observed 
variable $\delta x_n$ through the relation  
$\xi_n = \delta x_n/\bra \delta x_n^2 \ket^{1/2}$. 
The tail part and the center part of PDFs are connected at 
$\xi_n^*$ under the conditions that they have a common value and 
a common log-slope. The point $\xi_n^*$ has the characteristics that 
the dependence of PDF on $n$ is minimum for large $n$. We will see 
that $\xi_n^* \sim 1$ through the analyses of experiments. 
The tail part (\ref{PDF tl}) is determined by 
(\ref{Tsallis prob density}) with the translation of variable 
given by the first equation of (\ref{trans vari}). 
Note that the formulae (\ref{PDF tl}) and (\ref{PDF cra}) or (\ref{PDF crb}) 
are unified 
in the sense that it provides the PDFs of velocity fluctuations and 
of velocity derivatives with $\phi = 1$, the PDFs of pressure 
fluctuations and of fluid particle accelerations with $\phi = 2$, 
and the PDFs of energy transfer rates and of energy dissipation rates 
with $\phi = 3$. Note also that the energy dissipation rate is 
a variable taking only positive real values, whereas the others 
are variables taking both negative and positive real values. 
The PDFs for the latter variables are given by (\ref{PDF tl}) 
and (\ref{PDF cra}) with three parameters $q$, $n$ and $q'$. On the other hand, 
the PDF for the former variable is given by (\ref{PDF tl}) 
and (\ref{PDF crb}) with four parameters $q$, $n$, $q'$ and $\theta$.

These parameters are determined by the following procedure 
through a series of observed PDFs obtained by changing 
the distances of two measuring points, i.e.,
1) 
Start with a trial value $\mu$ (and also with trial values 
$q'$ and/or  $\theta$) to fit one of the observed PDFs with the 
tail part PDF given by (\ref{PDF tl}). Note that the values of 
parameters $\alpha_0$, $X$ and $q$ are determined as functions of 
$\mu$, self-consistently.
2) 
Once one has an appropriate $\mu$ value, other observed PDFs 
in the series can be fit with the correct increment $\it\Delta n$.
3) 
After getting the parameters $\mu$ (or equivalently $q$) and $n$ 
with trial values $q'$ and/or $\theta$, one can adjust better 
values for $q'$ and/or $\theta$ by fitting the center part of 
the observed PDFs with the formula (\ref{PDF cra}) or (\ref{PDF crb}).
4) 
Repeating the above process 1) $\sim$ 3), one can obtain the best 
set of parameters.

With the PDFs (\ref{PDF tl}) and (\ref{PDF cra}) or (\ref{PDF crb}), 
the $m$th moment of the 
quantity $\vert x_n \vert$ is given by
\be
\left\bra \vert x_n \vert^m \right\ket =
k \gamma^{(n)}_{\phi,m}
+ \left(1-k\gamma^{(n)}_{\phi,0} \right)\ 
a_{\phi m}\ \delta_n^{\zeta_{\phi m}}
\label{mth moment}
\ee
where $k = 2$ for the variables with the range ($-\infty$, $\infty$) 
and $k = 1$ for (0, $\infty$),
$
a_{3\phi m} = [ 2/\sqrt{C_{\phi m}} ( 1+ \sqrt{C_{\phi m}}) ]^{1/2}
\label{a_m}
$
and 
\be
\gamma^{(n)}_{\phi,m} = \int_{0}^{\infty} dx_n 
\vert x_n \vert^m [\Pi_\phi^{(n)}(x_n) - \Pi_{\phi,S}^{(n)}(x_n)].
\label{gamma}
\ee
Since the scaling exponents $\zeta_\bq$ is related to 
the generalized dimension $D_\bq$ by \cite{Meneveau87b}
\be
\zeta_{3\bq} = 1-(1-\bq)D_\bq, \quad (-\infty<\bq<\infty) 
\label{scaling exp 3qb}
\ee
one can determine the generalized dimension of the system 
through (\ref{mth moment}).

\section{Analyses of Simulations and Experiments\label{sec simulations}}

In Fig.~4 of \cite{AA7} and in Fig.~1 of \cite{AA8}, the PDFs of 
transverse velocity fluctuations
and of transverse velocity derivatives measured in the 
DNS on $2048^3$ mesh size \cite{Gotoh02} at $\mbox{R}_\lambda = 380$ 
are analyzed by the present theoretical 
PDFs (\ref{PDF tl}) and (\ref{PDF cra}) with $\phi = 1$ for velocity 
fluctuations and for velocity derivatives, 
and are plotted on (a) log and (b) linear scales. The observed PDFs 
are made symmetric by taking average of the data on the left and the 
right hand sides. The measuring distances, $r/\eta$, for the PDF of 
velocity fluctuations are, from the top to bottom in Fig.~4 of \cite{AA7}: 
2.38, 4.76, 9.52, 19.0, 38.1, 76.2, 152, 305, 609, 1220. 
We see from 
these values that the zooming increment of the consecutive PDFs is 
${\it\Delta} n = -1$. 
The spatial resolution of the DNS is $r_{\rm min}/\eta = 2.38$, and
the Kolmogorov length is $\eta = 2.58 \times 10^{-3}$.

\begin{figure}
\scalebox{0.5}{\includegraphics{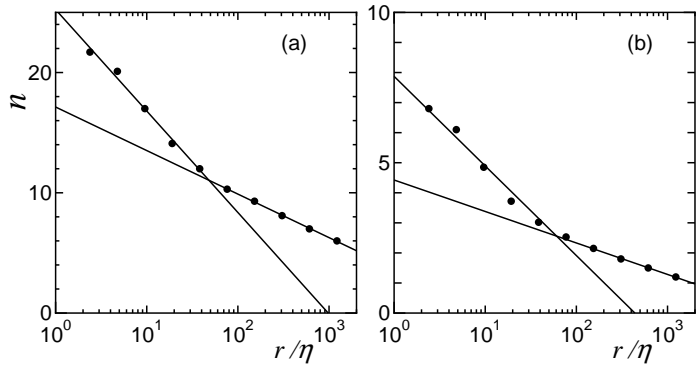}} 
\caption{Dependence of $n$ on $r/\eta$
\label{Dependence n}}
\end{figure}

Adopting $\mu = 0.326$ ($q = 0.543$) extracted form the analysis 
in Fig.~2 of \cite{AA7}, we have the parameters for the theoretical PDFs 
of velocity 
fluctuations in Fig.~4 of \cite{AA7}, from the top to bottom, 
($n$, $q'$) = (18.0, 1.90), (16.0, 1.80), (13.5, 1.85), (10.5, 1.75), 
(8.50, 1.65), (\underline{7.50}, 1.60), (\underline{6.50}, 1.50), 
(\underline{5.50}, 1.40), (\underline{4.50}, 1.30), 
(\underline{3.80}, 1.20). The dependence of $n$ on $r/\eta$ is plotted with 
closed circles in Fig.~\ref{Dependence n}(a). The lines are adjusted by 
the equations \cite{AA7}
\bea
n = -1.04 \log_2 r/\eta + 14.1 \quad \mbox{for} \quad r>r^{\rm T}, 
\label{n-roeta experimental Gotoh t large}\\
n = -2.39 \log_2 r/\eta + 21.1 \quad \mbox{for} \quad r<r^{\rm T}.
\label{n-roeta experimental Gotoh t small}
\eea
The parameters $n$ given above with underlines are contributing the points 
in Fig.~\ref{Dependence n}(a) adjusted by (\ref{n-roeta experimental Gotoh t large}), 
whereas those without underlines are 
contributing the points in the figure adjusted by 
(\ref{n-roeta experimental Gotoh t small}). 
The crossover occurs at $r^{\rm T}/\eta = 36.6$, which is close to the 
Taylor micro scale $\lambda/\eta = 38.3$ \cite{Gotoh02} of the system. 
The fact that (\ref{n-roeta experimental Gotoh t large}) provides us 
with the correct increment 
${\it\Delta} n = -1$ indicates that the scaling exponents in 
Fig.~2 of \cite{AA7} were extracted for $r> r^{\rm T}$ 
with the interpretation that 
the region is the inertial range for the DNS \cite{Gotoh02}.

Let us find out an appropriate $\mu$ value that gives the correct 
increment ${\it\Delta} n = -1$ for the points represented by 
(\ref{n-roeta experimental Gotoh t small}). In Fig.~\ref{Dependence n}(b), 
plotted is the dependence of 
$n$ on $r/\eta$ extracted with the appropriate $\mu$ value, 
$\mu = 0.850$ ($q = 0.882$), for the region $r < r^{\rm T}$ derived 
in accordance with the process 1) $\sim$ 4) given above. 
Then, we found the parameters for the theoretical PDFs of velocity 
fluctuations with this revised $\mu$ value, from the top 
to bottom, to be ($n$, $q'$) = (\underline{6.80}, 1.91), (\underline{6.10}, 1.90), 
(\underline{4.85}, 1.87), (\underline{3.72}, 1.78), (\underline{3.02}, 1.67), (2.53, 1.61), 
(2.15, 1.60), (1.80, 1.50), (1.50, 1.40), (1.20, 1.30). 
The lines given in Fig.~\ref{Dependence n}(b) are adjusted by the equations
\bea
n = -0.358 \log_2 r/\eta + 4.82 \quad \mbox{for} \quad r>r^{\rm T},
\label{n-roeta experimental Gotoh t large b} \\
n = -0.995 \log_2 r/\eta + 8.16 \quad \mbox{for} \quad r<r^{\rm T}.
\label{n-roeta experimental Gotoh t small b}
\eea
The parameters $n$ given with underlines are contributing the points in 
Fig.~\ref{Dependence n}(b) adjusted by 
(\ref{n-roeta experimental Gotoh t small b}), whereas those without 
underlines are contributing the points in the figure adjusted by 
(\ref{n-roeta experimental Gotoh t large b}). The crossover now occurs at 
$r^{\rm T}/\eta = 37.7$. 
Note that PDFs with the revised $\mu$ value are almost the 
same as those given in Fig.~1 of \cite{AA8} with the original $\mu$ value, 
i.e. it is impossible to see the difference in their appearance. 
This investigation indicates that there exists another range for 
$r < r^{\rm T}$ representing a scale invariance different from the inertial range. 
A detailed investigation of the range is one of the attractive future 
problems.



\begin{figure}
\scalebox{0.4}{\includegraphics{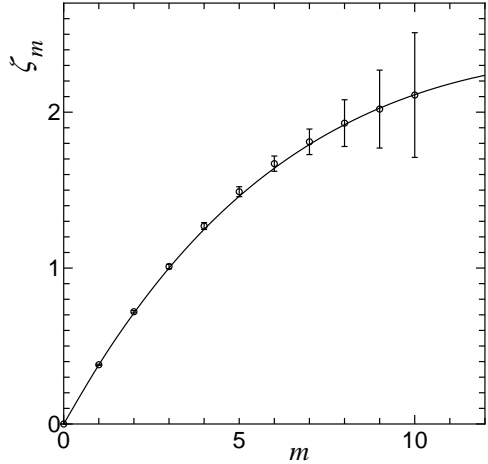}}
\caption{Scaling exponents
\label{scaling exponent Mouri}
}
\end{figure}

In Fig.~\ref{scaling exponent Mouri}, the transverse scaling exponents 
based on PDFs of velocity 
fluctuations extracted from the time series data obtained in a 
wind tunnel \cite{Mouri04,Mouri05} at $\mbox{R}_\lambda = 1054$ 
(closed circle) are analyzed by (\ref{zeta}) 
with $\mu = 0.357$ (solid line).

\begin{figure}
\scalebox{0.7}{\includegraphics{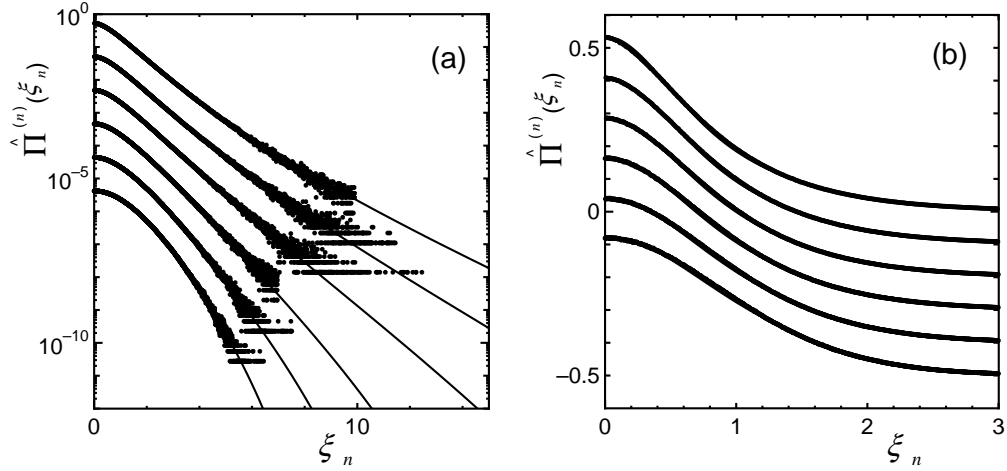}}
\caption{PDFs of Velocity Fluctuations
\label{PDF of velocity fluctuation}
}
\end{figure}

\begin{figure}
\scalebox{0.7}{\includegraphics{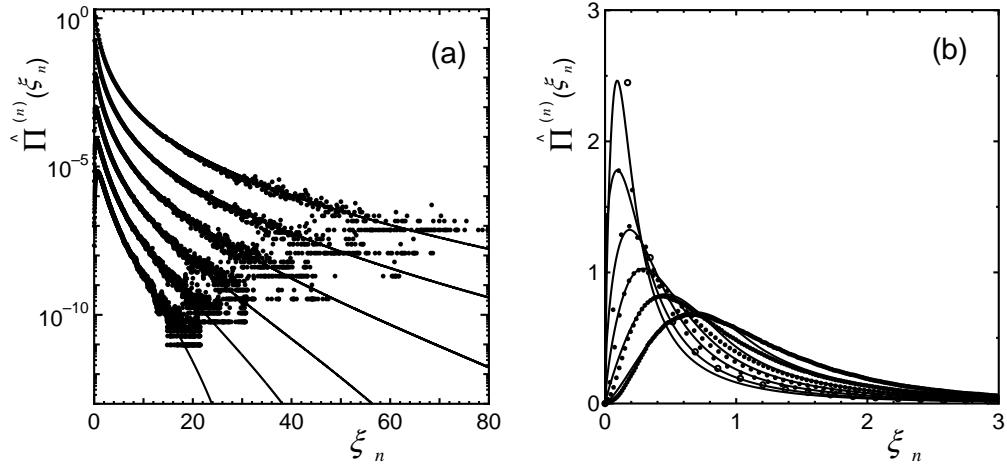}}
\caption{PDFs of Energy Dissipation Rates
\label{PDF of energy dissipation rate}
}
\end{figure}

In Figs.~\ref{PDF of velocity fluctuation} and \ref{PDF of energy dissipation rate}, 
displayed are PDFs of velocity fluctuations and of 
energy dissipation rates, respectively, extracted from the time series 
data \cite{Mouri04,Mouri05} with the help of Taylor's frozen hypothesis 
on (a) log and (b) linear scale. The distances $r/\eta$ between two 
measuring points for PDFs of velocity fluctuations and the lengths 
$r/\eta$ of the region in which energy dissipation rates are averaged 
to produce its PDF are, from the top to 
bottom: 28.2, 56.4, 113, 226, 451, 903. 
We see from these values that the increment 
of the consecutive PDFs is ${\it\Delta} n = -1$. The average wind velocity 
in the wind tunnel is 16~m/sec. The estimated inertial range is the region 
$44.7 < r/\eta  < 163$. 
The spatial resolution is 
$r_{\rm min}/\eta = 14.1$, and the Kolmogorov length is estimated as 
$\eta = 1.23 \times 10^{-2}$~cm. For the theoretical PDFs of velocity fluctuations, 
$\mu = 0.380$ ($q = 0.610$), 
($n$, $q'$) = 
(10.1, 1.78), (9.30, 1.70), (8.30, 1.65), (7.10, 1.62), (5.90, 1.55), (4.90, 1.50), 
whereas for the PDFs of energy dissipation rates, 
$\mu = 0.350$ ($q = 0.574$), ($n$, $q'$, $\theta$) = (7.60, 1.66, 1.81), 
(6.60, 1.20, 1.91), (5.60, 1.50, 1.77), (4.70, 1.68, 1.63), (4.20, 1.90, 1.44), 
(3.56, 2.30, 1.10). 
The values $\mu$ both of velocity fluctuations and of energy 
dissipation rates have been extracted following the process 1) $\sim$ 4) 
by adjusting the zooming increment to 
be ${\it\Delta} n = -1$ in the analyses of PDFs. The dependence of 
$n$ on $r/\eta$ for velocity fluctuations and for energy dissipation 
rates are, respectively, adjusted by the equations
$
n = -1.00 \log_2 r/\eta + 15.0
$
and
$
n = -1.00 \log_2 r/\eta + 12.4
$.
Note that the scaling exponents with
the derived $\mu$ values explain the extracted one in 
Fig.~\ref{scaling exponent Mouri} within the error bars.

\begin{figure}
\scalebox{0.4}{\includegraphics{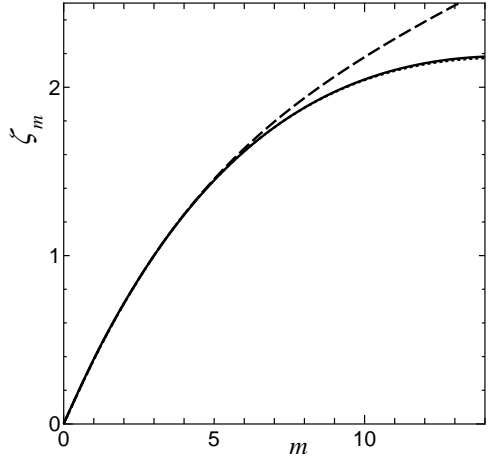}}
\caption{The Scaling exponents via PDF
\label{scaling exponent via PDF Mouri}
}
\end{figure}

In Fig.~\ref{scaling exponent via PDF Mouri}, 
solid line represents the scaling exponents $\zeta_m$ 
derived by the new route via observed PDFs of velocity fluctuations 
with the help of (\ref{mth moment}).
The left hand side of (\ref{mth moment}) 
is calculated with observed PDF 
data made up the lack of data for larger $x_n$ by the theoretical PDF 
$\Pi_{\phi,\mbox{S}}^{(n)}(x_n)$. Note that, within A\&A model, 
the difference between $\Pi_{\phi}^{(n)}(x_n)$ and 
$\Pi_{\phi,\mbox{S}}^{(n)}(x_n)$ for $x_n > x_n^*$ is neglected, 
where $x_n^*$ is the point corresponding to the connection point 
$\xi_n^*$. In the calculation of $\gamma_{\phi,m}^{(n)}$, only 
observed data is used for $\Pi_{\phi}^{(n)}(x_n)$, since the integral 
in (\ref{gamma}) stops at the upper limit $x_n^*$ which is about the order 
of the standard deviation.
The formula for the scaling exponents (\ref{zeta}) with 
$\mu = 0.380$ is shown in Fig.~\ref{scaling exponent via PDF Mouri} 
by dotted line which almost 
overlaps with solid line. Dashed line is the result obtained by the 
formula (\ref{mth moment}) {\it without} making up the lack of data for larger $x_n$ 
by the substitution of theoretical PDF $\Pi_{\phi,\mbox{S}}^{(n)}(x_n)$.

The PDFs of energy transfer rates and of energy dissipation rates measured 
in the DNS on $4096^3$ mesh size by Kaneda's group \cite{Aoyama05} at 
$\mbox{R}_\lambda = 1132$ are, successfully, analyzed by the present 
theoretical PDFs (\ref{PDF tl}) and (\ref{PDF cra}) with $\phi = 3$ for 
energy transfer rates and by PDFs (\ref{PDF tl}) and (\ref{PDF crb}) with 
$\phi = 3$ for energy dissipation rates. The observed PDFs of 
energy transfer rates are made symmetric by averaging the data on the 
left and the right hand sides. The measuring distances, $r/\eta$, 
for the PDFs both of transfer rates and of dissipation rates are 
13.7, 78.1, 449. The inertial range is estimated as 
$63 < r/\eta  < 224$, the Kolmogorov length as 
$\eta = 5.12 \times 10^{-4}$, and the Taylor micro scale as 
$\lambda/\eta = 66.2$ \cite{Aoyama05}. For the theoretical 
PDFs of energy transfer rates, $\mu = 0.320$ ($q = 0.534$), 
($n$, $q'$) = (9.00, 1.75), (6.50, 1.70), (3.80, 1.50), 
whereas for the PDFs of energy dissipation rates, 
$\mu = 0.350$ ($q = 0.574$), ($n$, $q'$, $\theta$) = 
(7.20, 1.60, 1.30), (4.80, 1.10, 1.70), (2.30, 1.10, 4.00). 
The dependence of $n$ on $r/\eta$ for energy transfer rates and energy 
dissipation rates are, respectively, adjusted by the equations
$
n = -1.04 \log_2 r/\eta + 12.9
$
and
$
n = -1.01 \log_2 r/\eta + 11.2
$.
Note that the value of $\mu$ has been extracted by adjusting the 
zooming increment to be ${\it\Delta} n = -2.5$ in the analyses of PDFs.

\section{Conclusions}

It is shown that the formulae of PDFs within A\&A model 
for velocity fluctuations, 
energy dissipation rates and energy transfer rates
explain, successfully, corresponding PDFs extracted 
from DNSs on 2048$^3$ and 4096$^3$ mesh sizes, 
and experiments performed in a wind tunnel. It is also shown that the 
formulae of PDFs with appropriate zooming increment $\itDelta n$ 
corresponding to experimental 
situation give a new route to obtain the scaling exponents of velocity 
structure function, including intermittency exponent.


Extracting $\zeta_{3\bq}$ for $-\infty < \bq < \infty$  through 
the formula $\bra \vert x_n \vert^{\bq} \ket$, given by 
(\ref{mth moment}) with $k = 1$, out of the PDFs of energy dissipation 
rates ($\phi = 3$), we can obtain the generalized dimension  
$D_{\bq}$ through (\ref{scaling exp 3qb}). It may be a direct proof 
of the multifractal distribution of singularities in real space. 
Results will be given elsewhere. Note that, for $\bq < 0$, 
contribution of the first term in (\ref{mth moment}) becomes conspicuous, 
i.e., we should subtract the contributions originated from the term 
violating the invariance of the scale transformation.


The authors would like to thank Dr.~K.~Yoshida for enlightening 
discussions. They are also grateful to Dr.~H.~Mouri and to 
Profs.~Y.~Kaneda and T.~Ishihara for their kindness to provide 
the authors with their data prior to publication.


\end{document}